\documentclass[a4paper,11pt]{article}
\usepackage{pos}

\usepackage{cleveref}

\newcommand{\ket}[1]{{|#1\rangle}}
\newcommand{\bra}[1]{{\langle #1|}}

\newcommand{\SU}{{\mathrm{SU}}}

\newcommand{\mdc}{\{\lambda_\ell\},\{\lambda_s\},\{\alpha_s\}}
\newcommand{\mdpc}{\{\tilde{\lambda}_\ell\},\{\lambda_s\},\{\tilde{\alpha}_s\}}
\newcommand{\chE}{\hat{{\cal E}}}
\newcommand{\chUp}{\hat{{\cal U}}_P}
\newcommand{\sing}{\mathbf{1}}
\newcommand{\doublet}{\mathbf{2}}
\newcommand{\trip}{\mathbf{3}}
\newcommand{\atrip}{\bar{\mathbf{3}}}

\renewcommand{\logo}{\relax}

\title{Confined and Deconfined Phases of Qubit Regularized Lattice Gauge Theories}

\author*[a]{Shailesh Chandrasekharan}

\affiliation[a]{Department of Physics, Duke University,\\
Box 90305, Duke University, Durham, NC 27708, USA}

\emailAdd{sch27@duke.edu}

\abstract{We construct simple qubit-regularized Hamiltonian lattice gauge theories formulated in the monomer--dimer--tensor-network (MDTN) basis that are free of sign problems in the pure gauge sector. These models naturally realize both confined and deconfined phases. Using classical Monte Carlo methods, we investigate the associated finite-temperature phase transitions and show that they exhibit the expected universality classes of conventional $\SU(N)$ lattice gauge theories in various spacetime dimensions. Furthermore, we argue that second-order quantum phase transitions separating the confined and deconfined phases are likely to exist. Such critical points would provide a nonperturbative route to defining continuum limits of qubit-regularized gauge theories, potentially allowing Yang--Mills theory and related continuum gauge theories to emerge from finite-dimensional lattice constructions.
}

\FullConference{The 42nd International Symposium on Lattice Field Theory (LATTICE2025)\\
2-8 November 2025\\
Tata Institute of Fundamental Research, Mumbai, India\\}

\begin{document}
\maketitle

\section{Introduction}
\label{sec1}

Qubit regularization provides a framework for studying continuum quantum field theories using infinitely many interacting quantum degrees of freedom, each residing in a finite-dimensional Hilbert space on a spatial lattice \cite{Chan2024}. While the idea of qubit regularization has attracted significant attention in the context of quantum computation \cite{Qsim2023}, its conceptual implications extend well beyond such applications. In particular, it sharpens our understanding of Wilson’s renormalization group (RG) flows and highlights the freedom in defining microscopic lattice gauge dynamics that nevertheless give rise to identical long-distance physics.

Here we focus on formulating a massive continuum quantum field theory (QFT), with particular emphasis on pure Yang--Mills theory. Following Wilson's RG framework, one begins with a lattice field theory tuned to a quantum critical point, where the long-distance physics is governed by an RG fixed point describing a scale-invariant continuum theory. This fixed point captures the ultraviolet (UV) behavior of the massive QFT of interest. In the case of Yang--Mills theory, the UV fixed point corresponds to the asymptotically free theory of weakly interacting gluons. The massive QFT in the infrared (IR) is then obtained by perturbing the critical lattice theory with a relevant operator that drives the RG flow away from the UV fixed point. In Yang--Mills theory, this relevant coupling is the gauge coupling. As the coupling flows toward the IR, the theory dynamically generates a mass gap, and the physical excitations in the spectrum are glueballs.

In this talk we argue that simple qubit-regularized lattice gauge theories can realize this scenario and potentially give rise to massive Yang--Mills theory in the continuum limit. In Sec.~2 we show that traditional lattice gauge theories can be reformulated in terms of color-electric flux variables. Using these variables, we construct a basis of the gauge-invariant Hilbert space that we refer to as the Monomer-Dimer-Tensor-Network (MDTN) basis. In Sec.~3 we introduce simple qubit-regularized models by truncating the MDTN basis to a small set of low-lying representations. In Sec.~4 we demonstrate that these models typically exhibit two distinct phases: a confined phase, in which static matter fields are confined, and a deconfined phase, in which they are not. We further argue that the finite-temperature confinement--deconfinement transitions fall into the same universality classes as those of traditional lattice gauge theories. In Sec.~5 we show that in simple quasi-one-dimensional models one can also identify second-order quantum critical points separating the confined and deconfined phases. These critical points provide a possible route to accessing a massive continuum QFT. Finally, in Sec.~6 we outline directions for future work.

\section{Monomer-Dimer-Tensor-Network (MDTN) basis}
\label{sec2}

Traditional lattice gauge theories are formulated as systems of quantum degrees of freedom living on the links $\ell$ of a spatial lattice, with each link variable taking values in the group $\SU(N)$. Equivalently, each link may be viewed as describing a quantum particle moving on the group manifold of $\SU(N)$. As explained in Ref.~\cite{Liu2022}, an orthonormal basis for the Hilbert space on a link $\ell$ is given by states $\ket{D^{\lambda_\ell}_{ij}}$, where $\lambda_\ell$ labels an irreducible representation (irrep) of $\SU(N)$, $i,j = 1,2,\dots,d_{\lambda_\ell}$ are representation indices, and $d_{\lambda_\ell}$ is the dimension of the irrep. The index $i$ transforms in the representation $\lambda_\ell$ associated with one end of the link, while $j$ transforms in the conjugate representation $\bar{\lambda}_\ell$ associated with the other end.

\begin{figure*}[t]
\centering \includegraphics[width=0.8\textwidth]{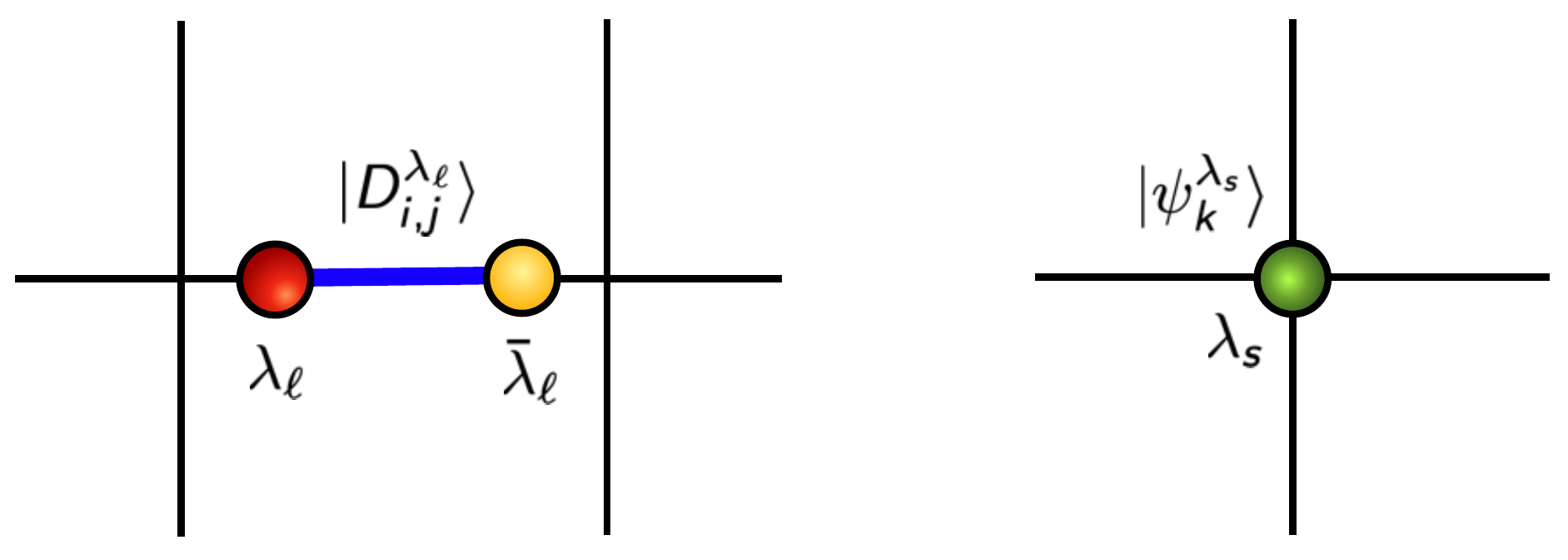}
\caption{Illustation of dimer-tensor (left) and monomer-tensors (right). Each circle stands for an index of either dimer-tensor or the monomer-tensor that transforms according the irrep of $\SU(N)$ shown in the figure.}
  \label{fig1}
\end{figure*}

In contrast, matter fields reside on lattice sites $s$. An orthonormal basis for the matter Hilbert space at a site $s$ can be written as $\ket{\psi^{\lambda_s}_k}$, where $\lambda_s$ is an irrep of $\SU(N)$ and $k = 1,2,\dots,d_{\lambda_s}$ is the corresponding representation index. The link states $\ket{D^{\lambda_\ell}_{ij}}$ may therefore be interpreted as {\em dimer tensors}, carrying two indices that transform under $\lambda_\ell$ and $\bar{\lambda}_\ell$ at the two ends of a link. Similarly, the site states $\ket{\psi^{\lambda_s}_k}$ may be viewed as {\em monomer tensors}, carrying a single index transforming under $\lambda_s$. A pictorial representation of these tensors is shown in \cref{fig1}.

\begin{figure}[t]
\centering \includegraphics[width=0.3\textwidth]{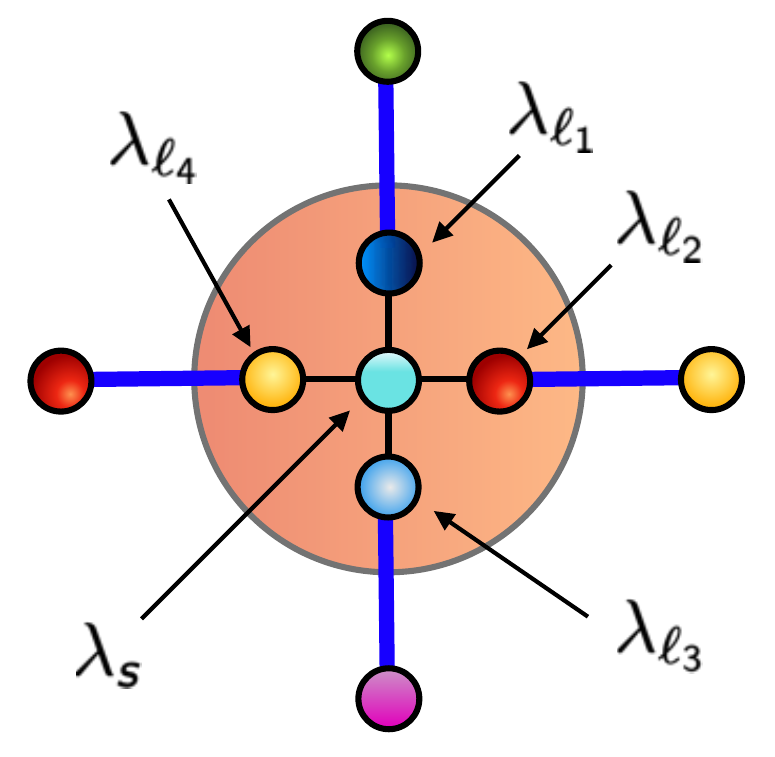}
\caption{Pictorial representation of the projection of dimer and monomer tensors onto the gauge-invariant physical subspace at a lattice site $s$. In the illustration, the local site Hilbert space before imposing Gauss's law is 
$
\mathcal{H}_s^{g} = \lambda_{\ell_1} \otimes \lambda_{\ell_2} \otimes \lambda_{\ell_3} \otimes \lambda_{\ell_4} \otimes \lambda_s \, ,
$
where the four link representations correspond to the links attached to the site. Imposing Gauss's law amounts to projecting onto the singlet (gauge-invariant) subspace of $\mathcal{H}_s^{g}$. We denote a basis of this singlet subspace by the index $\alpha_s = 1,2,\dots,\mathcal{D}_s$, where $\mathcal{D}_s$ is its dimension.
}
\label{fig2}
\end{figure}

A complete orthonormal basis of the full Hilbert space of a traditional lattice gauge theory is obtained by specifying a fixed set of dimer tensors on links, $\{\lambda_\ell\}$, and monomer tensors on sites, $\{\lambda_s\}$. However, this enlarged Hilbert space is not yet the physical Hilbert space, since physical states must be gauge invariant. The physical Hilbert space is obtained by projecting onto the gauge-invariant subspace through the imposition of Gauss's law. This projection can be implemented locally by constructing, at each lattice site $s$, the Hilbert space $\mathcal{H}_s^g$ on which gauge transformations act. This space is given by the tensor product of all irreps $\lambda_\ell$ associated with links attached to the site, together with the matter irrep $\lambda_s$ (if present). An illustration of $\mathcal{H}_s^g$ is shown in \cref{fig2}. The gauge-invariant subspace is the singlet sector of $\mathcal{H}_s^g$, obtained by appropriate tensor contractions (equivalently, by applying the fusion rules) among the indices meeting at site $s$. If the dimension of this singlet subspace is denoted by $\mathcal{D}_s$, we introduce an index $\alpha_s = 1,2,\dots,\mathcal{D}_s$ to label an orthonormal basis of gauge-invariant states at that site.

These observations imply that an orthonormal basis of the physical Hilbert space of a traditional $\SU(N)$ lattice gauge theory can be constructed from the data $\{\lambda_\ell\}$, $\{\lambda_s\}$, together with the singlet labels $\{\alpha_s\}$ specifying how the representations fuse to the trivial representation at each site. We denote the resulting orthonormal basis states by $\ket{\mdc}$ and refer to this construction as the monomer-dimer-tensor-network (MDTN) basis, since the physical states are ultimately described by a network of monomer and dimer tensors whose indices are contracted in such a way that gauge invariance is manifest~\cite{MDTN2025}.

\section{Qubit Regularized Lattice Gauge Theories}

Qubit-regularized lattice gauge theories can be constructed within the MDTN framework by restricting the allowed values of \( \lambda_\ell \) on the links appearing in the basis states \( \ket{\mdc} \). A simple qubit regularization scheme is obtained by restricting \( \lambda_\ell \) to the anti-symmetric irreps of \( \SU(N) \)~\cite{Liu2022}. We refer to this as the anti-symmetric qubit regularization (ASQR) scheme. For example, in the case of $\SU(2)$ we allow only $\lambda_\ell = \sing, \doublet$, while for $\SU(3)$ we choose $\lambda_\ell = \sing, \trip, \atrip$. More generally, for $\SU(N)$ we restrict $\lambda_\ell$ to the $N$ irreps (including the singlet) whose Young tableaux consist of a single column of boxes. This restriction renders the link Hilbert space finite dimensional and thus compatible with a qubit description.

While such truncated Hilbert spaces are widely used to define and study gauge theories on quantum computers, most work has focused on the traditional Kogut--Susskind Hamiltonian~\cite{KS1975}. In that formulation, matrix elements involve Clebsch--Gordan coefficients, which can introduce sign problems in certain the Monte Carlo approach. As a result, especially in higher dimensions, numerical studies are typically limited to small system sizes. It therefore remains unclear whether these simple qubit-regularized lattice gauge theories can support second-order quantum critical points and, if so, whether Yang--Mills theory can emerge from one of them in the continuum limit. The prevailing folklore is that the truncation must ultimately be removed in order to recover the true continuum theory.

\begin{figure*}[t]
\begin{center}
\includegraphics[width=0.9\textwidth]{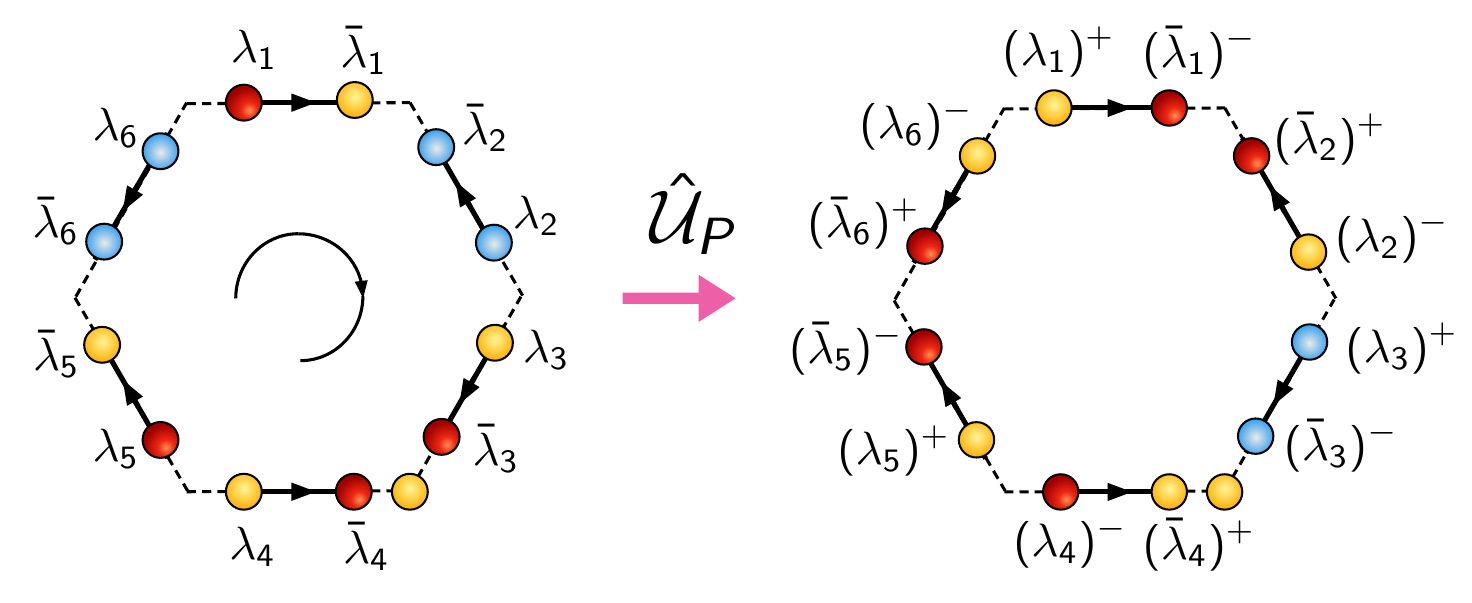}
\end{center} 
\caption{Action of the operator $\chUp$ on a plaquette in the $\SU(3)$ qubit-regularized gauge theory. The blue, red, and yellow circles represent the irreps $\sing$, $\trip$, and $\atrip$, respectively.}
\label{fig3}
\end{figure*}

To explore qubit regularization more systematically and to access larger system sizes---including with classical Monte Carlo methods---we proposed in Ref.~\cite{MDTN2025} a simpler class of Hamiltonians that can be written generically as
\begin{align}
H(\chE) = \sum_\ell \chE_\ell 
- \delta \sum_P \left( \chUp + \chUp^\dagger \right),
\label{eq:QRH}
\end{align}
where the sum over $\ell$ runs over links of the lattice and the sum over $P$ runs over plaquettes. The operator $\chE_\ell$ is diagonal in the MDTN basis and satisfies
\begin{align}
\chE_\ell \ket{\mdc} = \left(1 - \delta_{\lambda_\ell,1}\right) \ket{\mdc}.
\end{align}
In contrast, the plaquette operator $\chUp$ and its Hermitian conjugate are off-diagonal: they change the irreps $\{\lambda_\ell\}$ to $\{\tilde{\lambda}_\ell\}$ only on the links surrounding a given plaquette, while preserving gauge invariance. One of the simplest choices is
\begin{align}
\chUp 
= 
\sum_{\{\alpha_s\}, \{\tilde{\alpha}_s\}} 
\ket{\mdpc}\bra{\mdc},
\end{align}
where we assume that $\{\tilde{\lambda}_\ell\}$ and $\{\lambda_\ell\}$ differ only on the links belonging to the plaquette. An example of how the irreps change under this action in $\SU(3)$ is shown in \cref{fig3}. For further details of the construction, we refer the reader to Ref.~\cite{MDTN2025}.

\section{Confined and Deconfined Phases}

It is straightforward to argue that the qubit-regularized Hamiltonian in \cref{eq:QRH} describes a confined phase when $\delta = 0$. In this limit, the ground state has all link irreps equal to the singlet (i.e., $\lambda_\ell = \sing$ for all $\ell$). Introducing static matter fields at two distinct lattice sites requires the creation of a string of non-singlet irreps connecting the two sites. The energy of such a configuration grows proportionally with the separation between the sites, which is the hallmark of confinement.

\begin{figure*}[t]
\begin{center}
\includegraphics[width=0.5\textwidth]{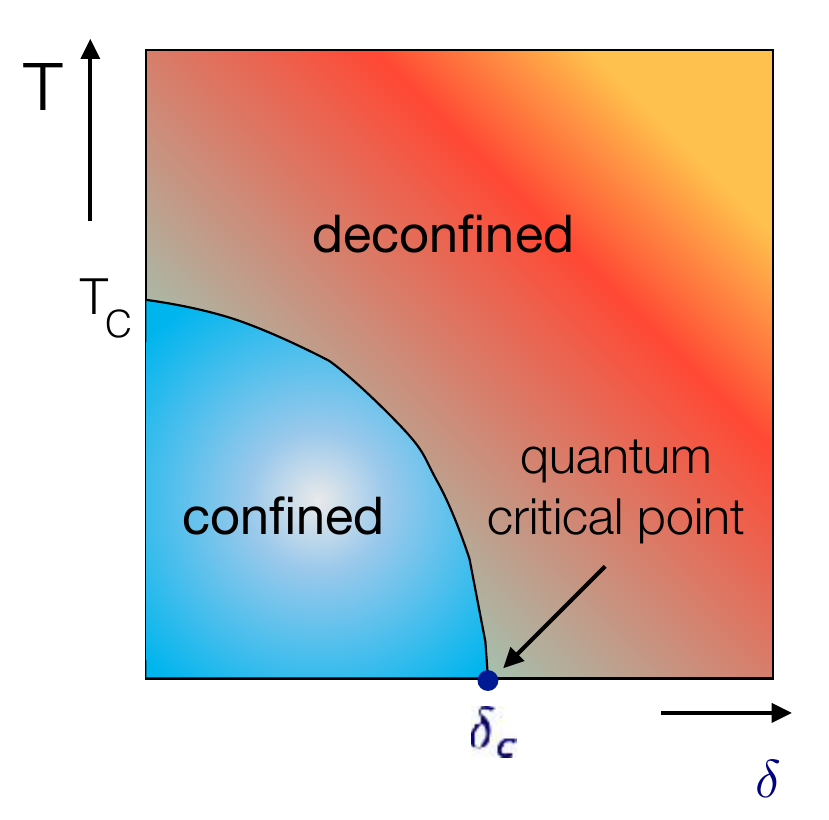} 
\end{center} 
\caption{A simple conjectured finite-temperature phase diagram of the qubit-regularized model described by the Hamiltonian in \cref{eq:QRH}. A massive continuum QFT may emerge from an appropriate relevant perturbation near the quantum critical point.}
\label{fig4}
\end{figure*}

At the opposite extreme, when $\delta \to \infty$, non-singlet irreps are no longer energetically suppressed. In this regime, introducing static matter fields at two distant sites does not require a linearly confining string, and the associated flux can spread out and fluctuate freely. This corresponds to a deconfined phase. Deconfinement also occurs naturally at sufficiently high temperatures, where thermal fluctuations allow string-like excitations to proliferate. These arguments suggest a simple finite-temperature phase diagram in the $T$--$\delta$ plane, as shown in \cref{fig4}.

Since the Hamiltonian in \cref{eq:QRH} does not suffer from sign problems, we can study its equilibrium thermodynamics using Monte Carlo methods on classical computers. While exploring the model at small temperatures as a function of $\delta$ is computationally challenging, the $\delta = 0$ limit can be studied efficiently as a function of the inverse temperature $\beta$. This allows us to investigate whether the qubit-regularized theory defined by \cref{eq:QRH} reproduces the universal physics of the finite-temperature confinement--deconfinement transitions of traditional $\SU(N)$ lattice gauge theories.

In $d$ spatial dimensions, these classical transitions are known to follow the order--disorder physics of $Z_N$ spin models: the low-temperature confined phase corresponds to the disordered phase of the spin model, while the high-temperature deconfined phase corresponds to the ordered phase~\cite{SY1982}.

\begin{figure*}[t]
\begin{center}
\includegraphics[width=0.95\textwidth]{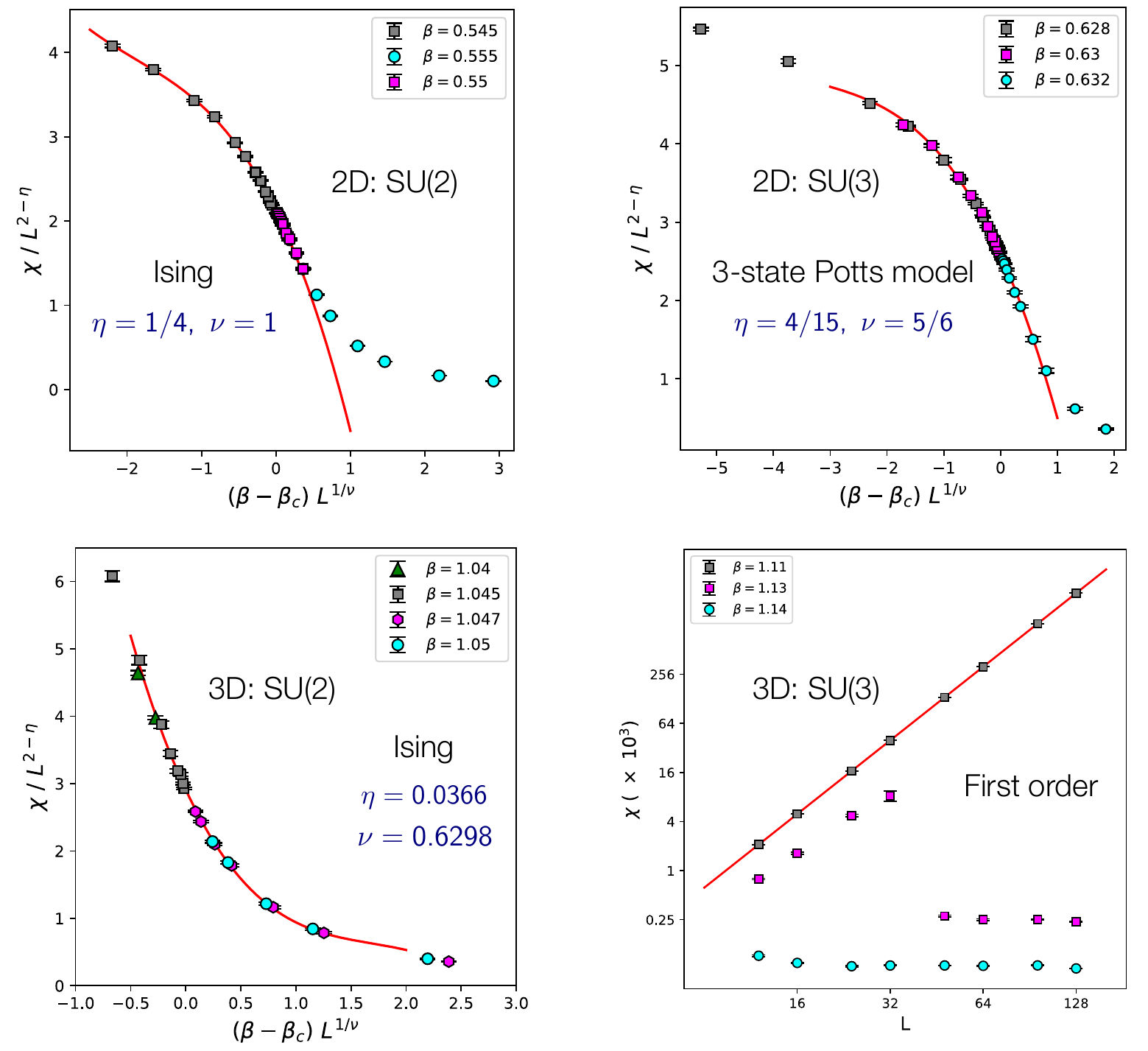} 
\end{center} 
\caption{Finite temperature critical behavior of \cref{eq:QRH} when $\delta=0$ with $\SU(2)$ and $\SU(3)$ gauge symmetries. The results shown are obtained using Monte Carlo methods in two (top row) and three (bottom row) spatial dimensions.}
\label{fig5}
\end{figure*}

To study the finite-temperature confinement--deconfinement transition, we focus on the pure gauge theory with $\lambda_s = 1$ on all lattice sites. Let $Z$ denote the partition function of this system. We then consider two lattice sites, $x$ and $y$, where heavy static matter fields are introduced so that $\lambda_x$ and $\lambda_y$ transform in either the fundamental or anti-fundamental irrep of $\SU(N)$. Let the corresponding partition function be denoted by $Z^{(x,y)}$. We define the susceptibility
\begin{align}
\chi = \frac{1}{L^d} \sum_{x,y} \frac{Z^{(x,y)}}{Z},
\end{align}
where $L^d$ is the spatial lattice volume.

In the thermodynamic limit, $\chi$ is expected to approach a constant in the confined phase and to scale as $L^d$ in the deconfined phase. If the transition at $\beta_c$ is second order, we expect the finite-size scaling form
\begin{align}
\chi L^{2-\eta} = f\!\left((\beta - \beta_c) L^{1/\nu}\right),
\end{align}
where $\nu$ and $\eta$ are the usual critical exponents.

In two spatial dimensions, traditional $\SU(N)$ lattice gauge theories with $N=2,3$ are expected to exhibit second-order transitions in the universality classes of the two-dimensional Ising and three-state Potts models, respectively. In three spatial dimensions, the $N=2$ transition belongs to the 3D Ising universality class, while the $N=3$ transition is expected to be first order.

We computed $\chi$ in the qubit-regularized theory as a function of lattice size $L$ for $d=2$ (honeycomb lattice) and $d=3$ (diamond lattice) using loop Monte Carlo algorithms~\cite{ALGO2003}. Our results are consistent with these expectations and are shown in \cref{fig5}. Further details can be found in Ref.~\cite{MDTN2025}.

\section{Quantum Critical Point and Continuum Limit}

Ultimately, the usefulness of qubit regularization for the study of Yang--Mills theory and related theories such as QCD depends crucially on whether a continuum limit can be reached in which these theories emerge. Such a limit can exist only if the qubit-regularized theory admits a relativistic second-order quantum critical point. Our conjectured phase diagram in \cref{fig4} suggests that such a critical point may occur at $\delta_c$, separating the confined and deconfined phases. 

In $d=3$ spatial dimensions, the continuum limit emerging at such a critical point must either reproduce Yang--Mills theory or define some other interacting $\SU(3)$ gauge theory. While the latter possibility may not be immediately relevant experimentally, it would nevertheless be theoretically exciting, as it would demonstrate that new types of continuum gauge theories can arise from simple qubit-regularized constructions.

\begin{figure*}[t]
\begin{center}
\includegraphics[width=0.95\textwidth]{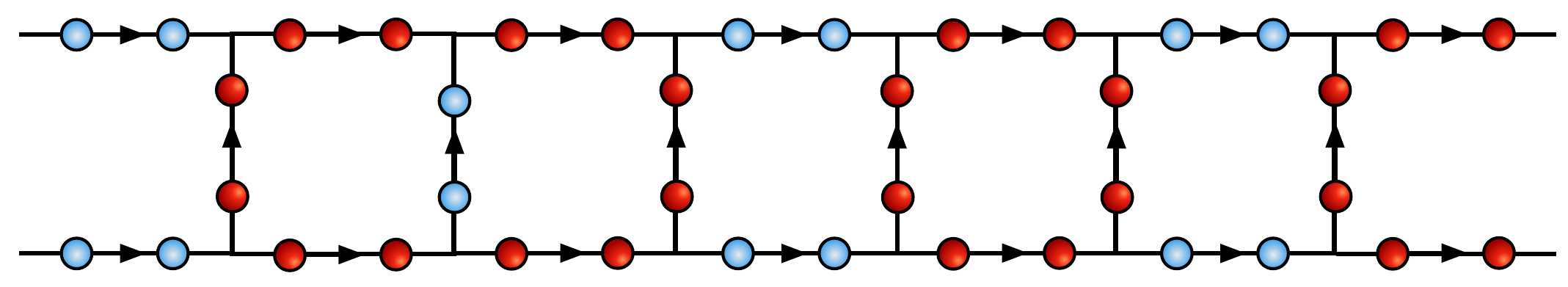} 
\end{center} 
\caption{An MDTN basis state of the $\SU(2)$ plaquette chain. The blue circles represent the irrep $\sing$, while the red circles represent the irrep $\doublet$. The configuration shown belongs to the even Hilbert-space sector, in which the top and bottom chain links on each plaquette carry the same irrep.}
\label{fig6}
\end{figure*}

Since studying \cref{eq:QRH} directly in $d=3$ is computationally challenging, we first explore the existence of quantum critical points and continuum limits in lower dimensions. A particularly simple system is provided by $\SU(N)$ plaquette chains, which are quasi-one-dimensional systems with nontrivial dynamics. An MDTN basis state on a plaquette chain for the $N=2$ case is shown in \cref{fig6}. We consider this case further to investigate if a continuum limit can exist. For this purpose we modify \cref{eq:QRH} by distinguishing between chain links $\ell_c$ and rung links $\ell_r$ that connect neighboring sites along the chain. The Hamiltonian becomes
\begin{align}
H \;=\; \kappa_c \sum_{\ell_c} \hat{\cal E}_{\ell_c}
+ \kappa_r \sum_{\ell_r} \hat{\cal E}_{\ell_r}
- \delta \sum_{P} \hat{\cal U}_P,
\label{eq:Hchain}
\end{align}
where $\kappa_c$ and $\kappa_r$ control the electric energy on the chain and rung links, respectively.

As explained in Ref.~\cite{Siew2025}, the quantum system described by \cref{eq:Hchain} naturally splits into two topologically distinct sectors. In the even sector, the two chain links on each plaquette are both either $\sing$ or $\doublet$, while in the odd sector one link is $\sing$ and the other is $\doublet$. In the even sector, the model can be mapped onto a transverse-field Ising model (TFIM) in a uniform magnetic field, with Hamiltonian
\begin{align}
H^E \;=\;
\sum_i \Big(
\kappa_c + \frac{\kappa_r}{2}
- \frac{\kappa_r}{2} \sigma_i^z \sigma_{i+1}^z
- \kappa_c \sigma_i^z
- \delta \sigma_i^x
\Big).
\label{eq:HIsing-E}
\end{align}

We note that when $\kappa_r = 2$, $\delta = 1$, and $\kappa_c = 0$, the TFIM is critical, and the lattice model is described in the continuum by a conformal field theory (CFT) of free massless Majorana fermions. As shown by Zamolodchikov, turning on $\kappa_c \neq 0$ generates a mass gap, and the resulting infrared theory is the $E_8$-invariant massive continuum QFT. Thus, in the $\SU(2)$ plaquette chain, our ideal scenario is realized explicitly: a qubit-regularized lattice gauge theory exhibits an asymptotically free ultraviolet fixed point and a massive relativistic infrared spectrum.

We can further compute universal mass ratios of the relativistic massive excitations, which may be interpreted as quasi-one-dimensional analogs of glueballs. To determine the string tension between static matter fields, \cref{eq:HIsing-E} must be modified to incorporate heavy charges at lattice sites. Further details of these mass ratios and string tension calculations can be found in Ref.~\cite{Siew2025}.

Although the above discussion focused on the $N=2$ case, the construction extends naturally to $N=3$. In that case, the qubit-regularized gauge theory maps to the three-state Potts model in a uniform magnetic field. The qualitative features found for $N=2$ persist for $\SU(3)$. Since these systems are one-dimensional, we employ density matrix renormalization group (DMRG) methods as the primary computational tool. Details of these calculations will be presented elsewhere.

The results for the plaquette chain provide hope that similar continuum limits may exist in higher spatial dimensions. To explore this possibility, however, efficient Monte Carlo methods for qubit-regularized gauge theories in higher dimensions are required. Since \cref{eq:QRH} is free of sign problems, such approaches should be feasible. We have recently developed a continuous-time path-integral Monte Carlo method for this purpose~\cite{Karna2025}, and we plan to apply it to higher-dimensional qubit-regularized lattice gauge theories in the near future.

\section{Conclusions}

In this work we have proposed a simple and conceptually transparent framework for constructing qubit-regularized lattice gauge theories using the monomer-dimer-tensor-network (MDTN) basis. By truncating the allowed link irreps to a finite set---such as the anti-symmetric representations of $\SU(N)$---we obtain gauge-invariant lattice models with finite-dimensional local Hilbert spaces that are naturally suited for both classical and quantum simulations. Importantly, by designing Hamiltonians that avoid sign problems, we are able to explore their equilibrium thermodynamics using large-scale Monte Carlo methods. Our results demonstrate that these qubit-regularized models exhibit confined and deconfined phases and reproduce the expected universality classes of finite-temperature confinement--deconfinement transitions in traditional $\SU(N)$ lattice gauge theories.

A central question is whether such truncated theories can admit relativistic second-order quantum critical points, thereby allowing a continuum limit in which Yang--Mills theory or other interacting gauge theories emerge. Our study of quasi-one-dimensional $\SU(N)$ plaquette chains provides strong evidence that this scenario can indeed be realized. In the $\SU(2)$ case, the model maps to the transverse-field Ising model and reproduces the well-known flow from a critical Majorana CFT to a massive $E_8$-invariant quantum field theory. This establishes explicitly that a qubit-regularized gauge theory can possess an asymptotically free ultraviolet fixed point and a relativistic massive infrared spectrum.

These results provide encouragement that similar continuum limits may exist in higher spatial dimensions. While establishing this in $d=3$ remains a major challenge, the absence of sign problems and the development of new Monte Carlo algorithms open a promising path forward. More broadly, our work suggests that qubit regularization is not merely a computational convenience, but a potentially powerful theoretical framework for discovering and classifying new continuum gauge theories emerging from finite-dimensional lattice systems.

\section*{Acknowledgments}

I thank T. Bhattacharya and R. X. Siew, for collaboration on several aspects of the work presented here. We acknowledge the use of AI assistance, specifically ChatGPT, in refining the language and clarity of this manuscript. This research was supported in part by the U.S. Department of Energy, Office of Science, Nuclear Physics program under Award No. DE-FG02-05ER41368.


\begin{thebibliography}{99}

\bibitem{Chan2024}
S. Chandrasekharan, {\em Qubit Regularization of Quantum Field Theories}, 
\href{https://pos.sissa.it/466/001}{PoS LATTICE2024 (2025) 001}[\href{https://arxiv.org/abs/2502.05716}{\tt{2502.05716}}].

\bibitem{Qsim2023}
C. W. Bauer, Z. Davoudi, N. Klco, and M. J. Savage, {\em Quantum simulation of fundamental particles and forces},
\href{https://doi.org/10.1038/s42254-023-00599-8}{Nature Rev. Phys. \textbf{5} (2023) 420} [\href{https://doi.org/10.48550/arXiv.2404.06298}{\tt{2404.06298}}].
%

\bibitem{DTh2004}
R. Brower, S. Chandrasekharan, S. Riederer, and U. J. Wiese, {\em D theory: Field quantization by dimensional reduction of discrete variables},
\href{https://doi.org/10.1016/j.nuclphysb.2004.06.007}
{Nucl. Phys. B \textbf{693} (2004) 149} [\href{https://doi.org/10.48550/arXiv.hep-lat/0309182}{\tt{hep-lat/0309182}}].

\bibitem{Liu2022}
H. Liu and S. Chandrasekharan, {\em Qubit regularization and qubit embedding algebras}, 
\href{https://doi.org/10.3390/sym14020305}
{Symmetry \textbf{14} (2022) 305} [\href{https://doi.org/10.48550/arXiv.2112.02090}{\tt{2112.02090}}].

\bibitem{MDTN2025}
S. Chandrasekharan, R.X. Siew, T. Bhattacharya, {\em Monomer-dimer tensor-network basis for qubit regularized gauge theories},
\href{https://doi.org/10.1103/gns9-l3gk}{Physical Review D 111, (2025) 114502}[\href{https://doi.org/10.48550/arXiv.2502.14175}{\tt{2502.14175}}].

\bibitem{KS1975}
J. B. Kogut and L. Susskind, {\em Hamiltonian formulation of Wilson’s lattice gauge theories}, 
\href{https://doi.org/10.1103/PhysRevD.11.395}
{Phys. Rev. D \textbf{11} (1975) 395}.

\bibitem{SY1982}
B. Svetitsky and L. G. Yaffe, {\em Critical behavior at finite temperature confinement transitions},
\href{https://doi.org/10.1016/0550-3213(82)90172-9}
{Nucl. Phys. B \textbf{210} (1982) 423}.

\bibitem{ALGO2003}
D. H. Adams and S. Chandrasekharan, {\em 
Chiral limit of strongly coupled lattice gauge theories}, 
\href{https://doi.org/10.1016/S0550-3213(03)00350-X}
{Nucl. Phys. B \textbf{662} (2003) 220} 
[\href{https://arxiv.org/abs/hep-lat/0303003}{\tt{hep-lat/0303003}}].

\bibitem{Siew2025}
R. X. Siew, S. Chandrasekharan and T. Bhattacharya, {\em Asymptotic-freedom and massive glueballs in a qubit-regularized SU(2) gauge theory}, [\href{https://doi.org/10.48550/arXiv.2512.11068}{\tt{2512.11068}}].

\bibitem{Zam1989}
A. B. Zamolodchikov, {\em Integrals of Motion and S-Matrix of the scaled {$T=T_c$} Ising Model with Magnetic Field}, \href{https://doi.org/10.1142/S0217751X8900176X}{Int. J. Mod. Phys. A\textbf{4} (1989) 4235}.

\bibitem{Karna2025}
A. Karna, W. Hansen, S. Chandrasekharan and R. Kaul, {\em Projected Density Matrix Sampling of Lattice Hamitonians}, [\href{
https://doi.org/10.48550/arXiv.2511.19209}{\tt{2511.19209}}].

\end{thebibliography}
\end{document}